\documentclass[manuscript]{acmart}

\AtBeginDocument{%
  }

\setcopyright{acmlicensed}
\copyrightyear{2018}
\acmYear{2018}
\acmDOI{XXXXXXX.XXXXXXX}

\acmConference[Conference acronym 'XX]{Make sure to enter the correct
  conference title from your rights confirmation emai}{June 03--05,
  2018}{Woodstock, NY}
\acmISBN{978-1-4503-XXXX-X/18/06}

\usepackage{hyperref}
\usepackage{cleveref}
\usepackage{xspace}
\usepackage{enumitem}
\usepackage{dblfloatfix} 
\usepackage{makecell} 

\begin{document}

\newcommand{\snp}{P\&S\xspace}
\newcommand{\psp}{PIPS\xspace}
\newcommand{\swc}{Smart Webcam Cover\xspace}
\newcommand{\cm}{Candid Mic\xspace}
\newcommand{\odrfid}{On-demand RFID\xspace}

\title[Physically-intuitive Privacy and Security]{Physically-intuitive Privacy and Security: A Design Paradigm for Building User Trust in Smart Sensing Environments }

\author{Youngwook Do}
\affiliation{%
\institution{Georgia Institute of Technology}
\department{School of Interactive Computing}
\city{Atlanta}
\state{GA}
\country{USA}
}
\email{youngwookdo@gatech.edu}

\author{Yuxi Wu}
\affiliation{%
\institution{Northeastern University}
\department{Khoury College of Computer Sciences}
\city{Boston}
\state{MA}
\country{USA}
}
\email{yuxiwu@gatech.edu}

\author{Gregory D. Abowd}
\affiliation{%
\institution{Northeastern University}
\department{Department of Electrical and Computer Engineering}
\city{Boston}
\state{MA}
\country{USA}
}
\email{g.abowd@northeastern.edu}

\author{Sauvik Das}
\affiliation{%
\institution{Carnegie Mellon University}
\department{Human-Computer Interaction Institute}
\city{Pittsburgh}
\state{PA}
\country{USA}
}\email{sauvik@cmu.edu}

\renewcommand{\shortauthors}{Do et al.}


\begin{abstract}
Sensor-based interactive systems---e.g., ``smart'' speakers, webcams, and RFID tags---allow us to embed computational functionality into physical environments.
They also expose users to real and perceived privacy risks: users know that device manufacturers, app developers, and malicious third parties want to collect and monetize their personal data, which fuels their mistrust of these systems even in the presence of privacy and security controls. 
We propose a new design paradigm, physically-intuitive privacy and security (PIPS), which aims to improve user trust by designing privacy and security controls that provide users with simple, physics-based conceptual models of their operation. PIPS consists of three principles: (1) direct physical manipulation of sensor state; (2) perceptible assurance of sensor state; and, (3) intent-aligned sensor (de)activation. We illustrate these principles through three case studies---Smart Webcam Cover, Powering for Privacy, and On-demand RFID---each of which has been shown to improve trust relative to existing sensor-based systems.
\end{abstract}

\ccsdesc[500]{Human-centered computing~Interaction design}
\ccsdesc[500]{Security and privacy}

\keywords{Usable Security and Privacy, Ubiquitous Computing}
\begin{teaserfigure}
  \includegraphics[width=\textwidth]{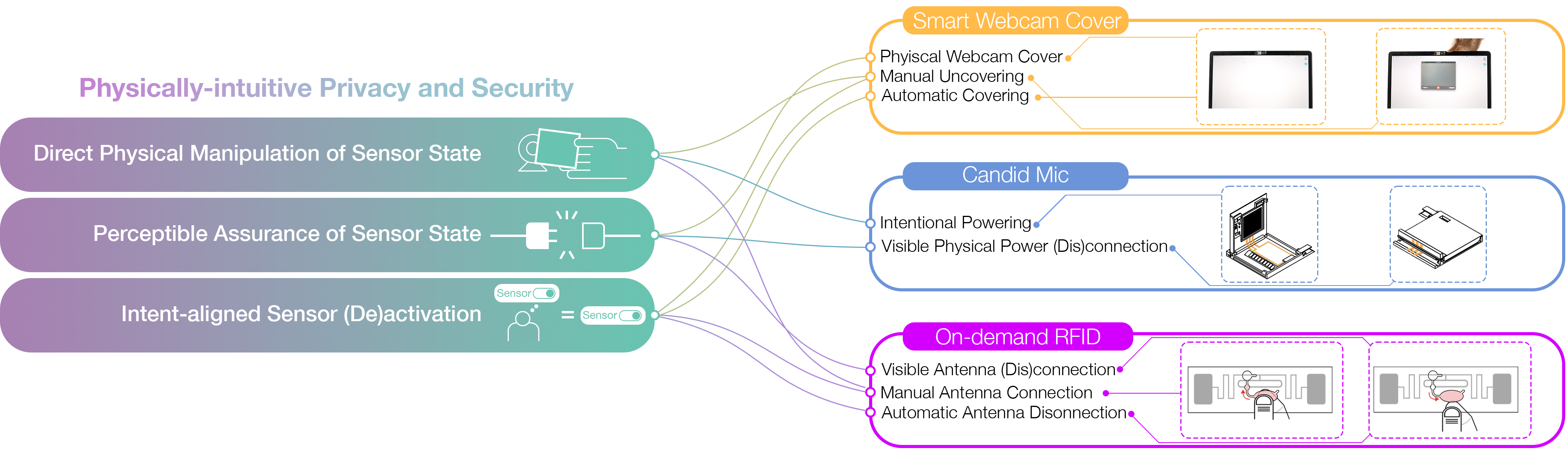}
  \caption{We introduce Physically-intuitive Privacy and Security (PIPS), a new design paradigm that tackles the challenge of improving user trust in sensor usage. PIPS takes advantage of users' physical intuition of perception to create \snp controls that are easy for users to understand and verify. PIPS is based on three principles: (1) direct physical manipulation of sensor state; (2) perceptible assurance of sensor state; and, (3) intent-aligned sensor (de)activation.}
  \label{fig:teaser}
\end{teaserfigure}

\maketitle

\section{Introduction}

The vision of Ubiquitous Computing (ubicomp), i.e., enabling seamless access to real-time, interactive computing anytime and anywhere, requires the proliferation of sensor-enabled devices into everyday environments \cite{weiser1991computer, weiser1999origins, abowd1998context}.
This broad integration of sensor-enabled devices into physical spaces allows for the creation of smart sensing  environments, objects, and things that can automatically infer users' social and environmental context and respond accordingly \cite{abowd1998context}.
However, a longstanding challenge that complicates this vision is the need for end-user privacy and security (\snp) \cite{langheinrich2001privacy}: if sensors are everywhere, it is imperative that users can understand and control, for example, what data of theirs is being collected, where it is going, how it is being used, and who has access to it \cite{langheinrich2001privacy,yao2019defending}.
A seemingly simple solution is for device manufacturers to provide users with privacy notice and controls---and indeed, prior art has explored many such solutions, ranging from privacy nutrition labels that improve user awareness (e.g., \cite{kelley2009nutrition}), to sophisticated permissioning systems that provide users with control (e.g., \cite{reeder2008expandable}).
Yet there remains an implicit distrust between device manufacturers and end-users \cite{lau2018alexa, smartspeakeruser_concerns, do2023powering, machuletz2018webcam} that continues to stymie this vision.

Within the broader context of surveillance capitalism \cite{zuboff2023age}, there is an inherent adversarial relationship between users who want to protect their personal data and device manufacturers who benefit from capturing as much personal data as possible.
Accordingly, many people think of device manufacturers---and the developers who build applications for those devices---as ``honest but curious'' adversaries who operate within the boundaries of the law and normative business practices but have strong incentives to collect data about users.
A good example is the widely held belief that advertisers and ad broker platforms can access smartphone microphones to ``eavesdrop'' on users' physical world conversations to better target them with ads \cite{kroger2019my}. 
Moreover, the presence of security vulnerabilities that allow for less honest third-party attackers to infiltrate and access these sensor-enabled systems---such as laptop webcams \cite{brocker2014iseeyou} and smart speakers \cite{defcon_2018}---further fuels this distrust.
Moreover, this distrust remains even in the presence of \snp notice and controls, such as webcam activation indications \cite{do2021smart} and the mute button on smart speaker microphones \cite{do2023powering}.

Don Norman helps explain why existing approaches to \snp control design leave room for this distrust \cite{norman2013design}:
\snp controls are rarely designed to provide users with a clear conceptual model of how they work.
Conceptual models are users' mental representation of how a system, object, or feature works [ibid].
Good conceptual models align with user expectations in a manner that makes it easy for users to predict the outcome of their actions, provide clear feedback and visibility, and are simple and consistent.
\snp controls often do not provide users with good conceptual models.
For the smartphone microphone example, users have little understanding of how permission systems work: when enabling the microphone permission for a particular app, there is limited, if any, indication of what has changed and how it has changed.
Thus, there remains room for distrust: even though the user may not have given the mobile app \textit{permission} to access their microphone, they cannot be sure that doing so has been made impossible. 
Prior work frames these poor conceptual models as a \textit{gulf} between how \snp controls actually work and users' understanding of how they work \cite{ahmad2020tangible}.




To bridge this gulf in the context of bystander privacy, Ahmad et al. introduced the concept of `tangible privacy', arguing that physically tangible, manipulable, and understandable \snp systems should mitigate bystanders' privacy concerns against primary users' sensor-enabled devices \cite{ahmad2020tangible,ahmad2022tangible}.
In a similar vein, Windl et al. have explored how tangible \snp controls can improve visitors' awareness of when sensor-enabled devices, deployed in a device owner's home, are activated \cite{windl2023investigating}. 



But what about when a user mistrusts their own devices?
How can we build smart sensing systems that users actually trust?
Fundamentally, the driving vision of ubicomp is about the \textit{physicalization} of computing: the elimination of seams between the worlds of atoms and bits.
This framing provides a helpful clue as to how we might better engender trust: perhaps we can improve trust by designing for people's \textit{physical} intuition, which has been developed over millions of years of evolution to make us acutely aware of physical risks and how to avoid them.
For example, we intuitively know that we can ``hide'' from a watchful gaze by breaking line-of-sight, and that we can lower our voice and \textit{whisper} if we want better control over who can hear us.
What if \snp controls for sensing systems were physically-intuitive in the same way so that there was little doubt about whether or not a sensor was capturing information?
To that end, in this paper, we describe and demonstrate a vision for \psp: an approach to designing sensing systems that engenders trust through physically intuitive design.

\psp relies on three key design principles: (i) direct physical manipulation of sensor state; (ii) perceptible assurance of sensor state; and (iii) intent-aligned sensor (de)activation. 
Direct physical manipulation of sensor state (e.g., covering a laptop webcam with an opaque piece of tape) helps users intuitively understand the mechanism by which they are enabling or disabling sensor capture. Perceptible assurance of sensor state (e.g., linking a sensor's power supply with an associated activation indicator) help users verify sensor state. Finally, intent-aligned sensor (de)activation (e.g., the automatic deactivation of sensor capture when users are no longer using a device) ensures that sensors can only capture data in line with user intent and expectation. 
When \snp controls are designed with some or all of these principles in mind, they improve user trust while still allowing users to access the functionality provided by sensing-enabled systems.

To illustrate these design principles in action, we will discuss three case study prototypes:  
Smart Webcam Cover \cite{do2021smart}, Candid Mic \cite{do2023powering}, and On-demand RFID \cite{do2025demand}. Each prototype provides \snp controls for a different sensing-based system, and embodies one or more of the aforementioned design principles. Each prototype has also been empirically validated to increase end-user trust.

In summary, our work introduces \psp---a new paradigm for designing physically-intuitive \snp controls for sensing-based systems that improves user trust. We envision that taking a \psp-inspired approach to designing \snp controls for sensing-based systems will, therefore, help overcome many of the end-user \snp concerns that encumber progress towards the vision of smart, responsive physical environments and the broader vision for ubicomp.

\section{Background and Definitions}
\label{section:scope}

``Security'', ``privacy'', ``trust'', and ``physical intuition'' all mean different things for different people and contexts. 
In this section, we define our usage of these terms throughout this paper and within our vision for \psp.

\subsection{What do we mean by ``security'' and ``privacy''?}

Security and privacy have been considered highly associated with each other and essential parts of social context regarding how to manage personal data \cite{bambauer2013privacy}. For example, Bambauer et al. define privacy as the framework concerning who has the right to access certain information, and security as the means to enable the framework [ibid]. In that regard, many parts of \snp are highly tied to access to each individual's physical space. (Indeed, Sylvester and Lohr's definition of privacy, which references a spectrum of personal data, includes an individual's personal physical space \cite{sylvester2005security}.) 
Ubicomp has expanded \snp in physical space to the online world, taking advantage of the advances in sensing capability to capture information regarding a user's physical space and enables the information to be wirelessly accessible. However, this accessibility puts users at risk of their physical space being remotely accessible by unauthorized actors \cite{medaglia2010overview}. 

When we refer to \snp in \textit{this} work, we mean \textit{\snp related to data collected by a user's sensor-enabled devices}. Specifically, we position privacy as a user's capability to ensure the alignment between what data is being collected by the sensor, and with whom the sensor is sharing the data, and what data users want collected and shared. Additionally, we frame security as a technical means to protect privacy --- ensuring the wrong actors cannot activate a sensor without a user's knowledge or consent. 

\subsection{What do we mean by ``ubiquitous sensors''?}

People have expressed \snp concerns against sensors in various contexts \cite{naeini2017privacy}. For example, past work has found that people may be worried about being surveilled without consent on public security cameras \cite{monahan2015right}.
However, while such sensors are embedded in users' everyday surroundings, in this paper, we focus on building trust for sensor-enabled devices that \textit{users own and operate themselves}.
For instance, people are concerned about smart speakers recording their activities when not in use~\cite{smartspeakeruser_concerns}.  In other words, devices not belonging to users are out of the scope of this paper, as users have no intention and/or control to use them, which makes it unnecessary to build trust in using them.

\subsection{What do we mean by ``trust''?}

Trust is a multi-faceted concept for which there is no singular definition \cite{mcknight2000trust}. McKnight and Chervany compiled research articles related to the definition of trust and surfaced a panoply of factors that comprise trust, including competence, benevolence, integrity, etc. [ibid] Blomqvist observes, however, that across these many factors, a common thread is that trust is often grounded in how much a counterparty will meet one's expectations in the future \cite{blomqvist1997many}. 
Similarly, O'Hara defines trust as an attribute of an individual that can be obtained by fulfilling what one promises \cite{o2012general}. 

As automation of system operations becomes more prevalent, negating the need for human intervention, human--computer trust is increasingly of interest to computing researchers \cite{madsen2000measuring}. Madsen et al. defined human--computer trust as how confident and willing a user is in following the decisions made by an intelligent computing system [ibid]. As existing \snp operations have become automated, how to improve perceived trust in \snp operations has become increasingly studied for sensor-enabled devices as a way to address \snp concerns \cite{ahmad2022tangible, seymour2023ignorance}.
Based on this prior work, we define trust in this paper as an end-users' confidence that the \snp operations of sensor-enabled devices work as they expect.

\subsection{What do we mean by ``physically-intuitive''?}
Leveraging people's knowledge of actions and constraints that make sense in the physical world, i.e., ``physical affordance'' \cite{norman2013design, ishii1997tangible}, can be an effective way of designing \snp operations that make sense to users.
However, designing \snp operations to be ``physically-inspired''---i.e.,  applying physical or tangible properties to a design, regardless of whether those properties are perceivable or understandable to users---is not sufficient for building user trust.  For example, inaudible sounds can be used to thwart a microphone recording \cite{chen2020wearable}, even if users cannot perceive the sounds. 
In this work, we use the term \textit{``physically-intuitive''} as a metonym for the design principles of \psp, which have been inspired by the concept of physical affordance. We refer to a design as being ``physically-intuitive'' if it leverages people's knowledge of the physical world to provide an intuitive understanding of the sensor state and its capture mechanism.

\subsection{Types of Data Control}
Various stages in the data management life cycle---data collection, transfer, storage, and processing, etc.---can impact end-users' privacy \cite{spiekermann2008engineering}. For example, in the data collection stage, sensor devices might collect user data without consent, exacerbating users' \snp concerns. In the data transfer and data storage stages, users may worry about unauthorized entities accessing their data, either through system access control policies or data breaches.
Users may also wonder about how their data is being analyzed and processed after the sensors record and transfer it to the cloud.
It may be challenging or infeasible for a user to have agency and control over the data transfer, storage, and processing stages, because they are typically back-end processes that users cannot access. To that end, in our paper, our focus is to design \snp operations that afford end-users agency in the front-end.   In other words, we focus on designing \psp controls that protect  against unwanted data collection. 


\subsection{Adversaries}


We envision \psp to be effective in building user trust that their ubiquitous sensors cannot be compromised by threat actors who aim to wirelessly and unobtrusively capture data. Note that we assume that these threat actors do not have physical access to end-users' sensors.

\subsection{Summary of Threat Model and Scope}
To summarize, in this paper, we outline a vision for \psp controls for ubiquitous sensors. We argue that this approach can help build user trust that these sensors only capture data in-line with their knowledge and consent, even in the presence of remote and unobtrusive adversaries: i.e., honest-but-curious device manufacturers and app developers, as well as less honest remote third-party attackers. These controls make the sensor state easy to perceive and the mechanism to allow or block capture easy to understand through simple physics-based interactions. In a future section, we will more concretely outline requirements for \psp controls.










\section{Case Studies}
\label{section:casestudies}
In this section, we consider, as case studies, three widely-deployed sensors---a laptop webcam, a smart speaker microphone, and a passive RFID tag. We chose these three cases by two dimensions: (1) perceptible assurance by physical barrier, and (2) accessibility to data-collecting devices. For example, webcams can be covered by a physical cover (e.g., tape), providing assurance that the cover prevents webcam recording. On the other hand, a microphone could still capture sound even if a user use a physical barrier (e.g., going to another room and closing its door) because sound could propagate through physical media, eroding, in turn, the assurance of a physical barrier to prevent data collection. Lastly, RFID sensing is imperceptible to users as electromagnetic signals. In addition, for passive RFID, users are often inaccessible to the RFID reader for control as the access infrastructure (e.g., gate access system). Even the users may not be aware of where the sensors are located. The detailed background for each sensor type will be further explained in the following subsections. We discuss the \snp concerns and threat model for each sensor type. Then, in the following section, we will discuss \psp design principles and how they might be applied to address \snp concerns against its associated threat model. 

\subsection{Case Study 1: Webcam}
This case study's focus is on the webcam of a user's own laptop device. This focus excludes the webcam of devices owned by others. 

\subsubsection{Background}
Webcams have become a point of privacy vulnerability as they have been widely deployed in private settings \cite{neustaedter2006blur}. As a visual cue to help users notice a webcam's activation, many laptops have been produced with its associated LED indicator. Despite this effort, people mistrust these LED indicators. Popular reports of law enforcement and/or malicious actors being able to manipulate and suppress laptop webcam LED indicators fuel this distrust \cite{koelle2018beyond}.

Accordingly, prior work has shown that many users take matters into their own hands by obstructing their webcams, when not in use, with a physical barrier (e.g., tape, sticky note, slider, etc.) \cite{machuletz2018webcam, HPSurvey71:online}. 
This crude method increases trust because users understand, through physical intuition, that when one places an opaque physical barrier in front of a camera or an eye, that object breaks line-of-sight \cite{koelle2018beyond}.  Moreover, it is easy for users to verify that the camera is blocked in a manner that no remote adversary can subvert.

However, manually obstructing a webcam is a cumbersome process that requires users to have a barrier on their person, and remember to put it back on every time they remove it for situations where they genuinely need to use their webcams. 
Unsurprisingly, people often forget to re-cover their webcams when no longer in use, which puts them back at risk of covertly being monitored \cite{do2021smart}. 

\subsubsection{Threat Model}
The target adversary of this case study is a malicious actor who can remotely manipulate the LED indicator associated with the webcam of a user's laptop. This threat model rules out situations where a malicious actor can physically access the user's space and device. The actor's goal is to surreptitiously record a user's physical space and/or a user's activities in the physical space.


\subsection{Case Study 2: Smart Speaker Microphone}
This case study is centered on a microphone embedded in a smart speaker device (e.g., an Amazon Echo). The sensor type we discuss in this case study preclude smart speakers owned by others. 

\subsubsection{Background}
Commodity smart speakers often have mute/unmute buttons for their microphones. End-users who do not want their conversations recorded can press the button to ``mute'' the microphone --- i.e., prevent the microphone from being able to actively ``listen'' in on any conversations. However, many people do not fully trust that the mute/unmute button fully prevents adversaries from capturing audio without users' knowledge or consent \cite{lau2018alexa}. Note that this distrust can exist even in the presence of mechanisms that do fully prevent microphones from capturing audio \cite{bezos_2021}. Users know that sensor manufacturers can benefit from collecting their personal data, and mute buttons do not make it clear how they are preventing capture, leaving from for distrust.

To address their distrust, many end-users simply power off their smart speakers when they want assurance that the device cannot ``listen'' \cite{lau2018alexa, jin2022exploring, chandrasekaran2021powercut}. Users do this even though it is inconvenient---they have to unplug the device tethered to wall outlet, as unplugging cannot be done wirelessly. Additionally, as the device would require rebooting upon their use, they have to wait until the re-activation, which could compromise the smart speaker's usability \cite{egelman2010please}.

\subsubsection{Threat Model}
The adversary in this case study can remotely access a user's smart speaker microphone and manipulate the microphone to record without a user's knowledge or consent. This threat model excludes the cases for a malicious actor to physically be present in a user's space and manipulate the user's device. The goal of the actor is to eavesdrop on the audio captured by a smart speaker microphone without the user's knowledge.

\subsection{Case Study 3: Passive RFID Tag}
This case study focuses on passive RFID tags that a user may own and carry on their person. Therefore, this focus disregards active RFID tags that are battery-powered. 

\subsubsection{Background}
Passive Radio Frequency Identification (RFID) technology has enabled numerous contactless interactions in everyday setups such as contactless payment with credit cards, key fob for door access, etc. This technology requires two components to run---a passive RFID tag and the tag reader. Specifically, as the tag is battery-free, the reader wirelessly activates the tag's data transfer so that the reader can receive the data. While the passive tag has a benefit of no need to recharge, this benefit conversely causes the vulnerability that the tag information could be unwittingly scanned as long as the tag reader is in vicinity. To that end, end-users would have to powerlessly give away their information stored in the tags if a malicious actor co-located in a physical space tries to contactlessly read the information. 

To address this risk, people use a physical RFID-blocking wallet designed to prevent any RFID signals from transceiving through the wallet \cite{koscher2009epc}. RFID-blocking wallets have metallic materials coated inside, which interferes with electromagnetic signals. While the wallets are supposed to block malicious actors' covert tag reading, Koscher et al. discovered that metal sleeves may not fully block RF signals [ibid]. This is critical as creating a discrepancy between how end-users expect the wallet to protect and what the wallet can protect, which could lead to eroding trust in using this physical protection for passive RFID tags. 

\subsubsection{Threat Model}
The adversary of this case study is a nefarious actor who is physically in the vicinity of a user who possesses RFID tags containing the user's sensitive information. The actor carries a passive RFID reader and covertly scans RFID tags' information without RFID tag owners' knowledge and consent. 
\section{Design Principles and Illustrative Prototypes}

In this section, we explain three \psp design principles and how those principles can be applied to designing \snp controls that address the concerns introduced in \Cref{section:casestudies}.

\subsection{\psp Design Principles}
Today, users' often reclaim agency over untrusted sensors through preventative physical actions, e.g., blocking and unblocking camera-enabled devices with a piece of paper \cite{machuletz2018webcam}, or pulling the plug on microphone-enabled ones \cite{lau2018alexa,jin2022exploring,sciuto2018hey,chandrasekaran2021powercut}.  However, whether they choose to take these actions or not, users face tradeoffs  \cite{zeng2017end,taylor2003most}.  If users opt \textit{not} to take these preventative physical actions, they must trust in software controls to maintain their privacy preferences, but prior art shows that users often mistrust software controls, believing, for example, that their microphones can eavesdrop on them at any moment \cite{smartspeakeruser_concerns} or that their webcams may be covertly accessed \cite{brocker2014iseeyou}. This mistrust is further amplified by the fact that software-based controls over sensors have been shown to be exploitable by threat actors (e.g., through covert manipulation of the LED indicator control associated with a webcam activation status. \cite{brocker2014iseeyou}) and can work inconsistently at times, as in the case of ``wake word'' detection controls for smart speakers \cite{dubois2020speakers,schonherr2020unacceptable,vaidya2015cocaine}. 


This disconnect, whether perceived or actual, between when users explicitly \textit{want} access to sensor-enabled functionality and when those sensors are, in fact, enabled and capturing data fuels the distrust many users harbor over sensor-enabled devices.
In this section, we describe three key characteristics of \psp that can help improve trust:

\begin{description}
   \item [Direct Physical Manipulation of Sensor State]
   Sensor controls should be understandable and analogous to physically-intuitive actions in everyday life, e.g., hiding things under opaque covers.
   \item [Perceptible Assurance of Sensor State] Capture state indicators should be noticeable and physically-guaranteed: e.g., if a sensor can only capture data through harvested energy, then the state indicator should use that same energy to indicate to users that the sensor is in ``capture'' mode.
   \item [Intent-aligned Sensor (De)activation] Sensors should be automatically deactivated in line with user use and expectation, and only manually activated through an intentional physical interaction.
\end{description}

\subsubsection{Direct Physical Manipulation of Sensor State}
One of the key elements of \psp is to let end-users physically manipulate \snp operations outside of sensor devices, instead of needing to blindly trust the operations inside of them.  Even when users should ostensibly have control over the \snp controls provided by sensors, e.g., if they own the sensor devices, the increasingly digitalized nature of this control means there is a gulf between how users \textit{expect} these controls  to work and how they actually work \cite{ahmad2020tangible}.  To narrow the gulf, the mechanism through which \psp controls enable or disable sensor capture should be physically intuitive and analogous to physical-world actions: e.g., drawing one's curtains at night. We draw on the concept of ``direct manipulation'', where a user directly performs operations and reviews the results instead of relying on the system explaining the operations or the results \cite{hutchins1985direct,ishii1997tangible,fitzmaurice1996graspable}, and propose \textbf{direct physical manipulation of sensor state} as a key design principle of \psp.

There are a few physical methods to thwart or interfere with sensor operation explored in past work. 
One is \textit{jamming} \cite{chandrasekaran2021powercut,chen2020wearable,sun2020alexa,truong2005preventing}. For example, if a person does not want to be recorded by a camera, they might try repeatedly turning a bright lamp on and off, such that any cameras cannot capture a usable video feed due to being jammed by the fluctuations in light exposure. With this method, however, people might still not be completely certain that the flashing lamp light can fully disrupt the camera feed; the light may partially jam the camera's CMOS sensor, allowing some parts of the recorded images to still be visible to the camera. Thus, jamming may not work as users intend, which still leaves a chasm between users' expectations and the actual effectiveness of the jamming operation. 

Another approach explored by prior art is to build physical analog signal filters that prevent raw data from being transferred to off-device storage for processing; these filters stand in contrast to software-based filters, which leave open the possibility of exploitation by third-party attackers for covert access to raw sensor data. PrivacyMic demonstrates a hardware design that leverages analog filters to allow only inaudible sound, which contains no sensitive conversation data and negates the need to pass raw data to any processor in the device \cite{iravantchi2021privacymic}. While these filters reduce the fidelity of data collected, they are not physically intuitive and thus are unlikely to improve trust: to many users who do not have knowledge of how electronic circuits work, the mechanism by which their data is being modified by these analog filters is non-obvious and non-intuitive. 


Instead, many people opt for simpler preventive actions that align with their physical intuition: e.g.,  covering a camera with an opaque object, such as a sheet of paper.  This direct physical manipulation is intuitive: by breaking line-of-sight, we understand naturally that we can no longer be seen. Moreover, the effects are easy to verify.  
Another preventive physical action that people take to interfere with or manipulate sensor operation is simply unpowering them by, e.g., removing their batteries or unplugging them \cite{ahmad2022tangible, chandrasekaran2021powercut, lau2018alexa, do2023powering}. Doing so incurs a utility cost, as accessing the functionality of the device requires users to re-power the devices prior to use.



\subsubsection{Perceptible Assurance of Sensor State}

Many sensors come with state indicators to indicate whether or not they are currently capturing information, but users may not trust these indicators if they are not physically tied to capture operations. For example, while many smart speakers have ``mute mode'' indicators to convey to users that they are not actively ``listening'', prior work has found that users do not always believe these indicators \cite{do2023powering}.
Indeed, users have little perceivable guarantee that ``mute mode'' in smart speakers is anything more than an LED strip coming active: after all, the LED strip has nothing to do with how the microphone captures audio data.
To address this gap between indication and operation, we propose \textbf{perceptible assurance of sensor state} as the second defining characteristic of \psp. 


Power is one way to confirm the state of capture in sensors.  As we noted in the previous principle, people unplug sensor devices from power to ensure that these sensors are not capable of unwanted capture. 
However, the mechanism through which power is cut should be visible and verifiable to end-users for it to be effective.
For example, consider a smart speaker mute button that kills power to the microphone sensor: if the process by which that power is killed is not intuitively understood or verifiable by the end-user, there is still room for distrust.
Beyond perceptibility, state changes should be \textit{physically guaranteed}.
For example, if a user \textit{perceives} that a sensor is in a specific capture state, that state should be guaranteed to be true---e.g., by linking the power source between the sensor and its use indicator.
Physically-guaranteed state change indicators, thus, must tether sensor capture state with indication of that state in a manner that is clear and verifiable by users.


\subsubsection{Intent-aligned Sensor (De)activation}
Human-in-the-Loop security systems must anticipate and account for user error in \snp operations \cite{cranor2008framework, sasse2001transforming}.
Sensor systems with \snp controls that require manual operation are no exception.
For instance, many users manually occlude laptop webcams with sticky notes, paper, and other adhesive covers: but they must remember to uncover their webcams when they actually need to use them, and then they remember to \textit{re-}cover their wecbams after they are done using them.
Unsurprisingly, the majority of manual webcam users report forgetting to recover their webcams \cite{do2021smart}.

In short, a system that relies on human memory for \snp operations both increases the burden on users and exacerbates their vulnerability to \snp threats.
Automated systems can reduce this burden, but do not necessarily breed trust: if the automation occurs outside of a user's conscious awareness, then they still require blind trust on the part of the user.  
To reduce user burden, exposure to threats, and build trust, we introduce \textbf{intent-aligned sensor (de)activation} to reduce reliance on user memory to deactivate sensors, and align sensor activation with intentional user interaction.
Intent-aligned sensor (de)activation requires two equally important components: automated and physically-guaranteed deactivation after an intentional physical interaction ends, and manual physical activation. 

Consider, for example, a sensor that must harvest power from a direct user interaction to begin capture (e.g., through the use of photovoltaics exposed to light). The sensor could only be \textit{activated} if a user interacts with it because that is how it draws power. Also, once a user ceases interacting with it, it must necessarily deactivate as the user is no longer supplying power. Manual activation coupled with physically-guaranteed deactivation ensures that the sensor capture state is always aligned with user expectations.  This way, a system can balance human-in-the-loop in \snp operations despite the benefits of automation \cite{edwards2008security,spiekermann2006technology}.

\subsection{Illustrative Prototypes}

To operationalize what we mean by \psp, we will highlight and analyze three illustrative research prototypes of sensors or sensor accessories that were developed to improve user trust through physics-inspired design. Recall that by ``improving trust'', we mean user's belief that data captured by the sensor aligns with their preferences and consent. 

For each of the three research prototypes we highlight --- \swc, \cm, and \odrfid --- we showcase how \psp concepts were employed to create trust-building controls and indicators. Each of these prototypes address the threat and trust challenges for the case study sensors we introduced before --- a webcam, a smart speaker microphone, and a passive RFID tag. \swc
illustrates how to use a physical barrier and intelligent automation to provide users' with perceptible assurance as to the state of data capture. \cm elucidates how to use intentional powering to ensure that a microphone can only ``listen'' when a user wants to be heard. Lastly, \odrfid demonstrates how ``off-by-default'' failsafe designs that work through intuitive, mechanical mechanisms can mitigate   \snp concerns.
We will further discuss how each prototype embodies each of the \psp design principles.

\begin{figure}
    \centering
    \includegraphics[width=1\columnwidth]{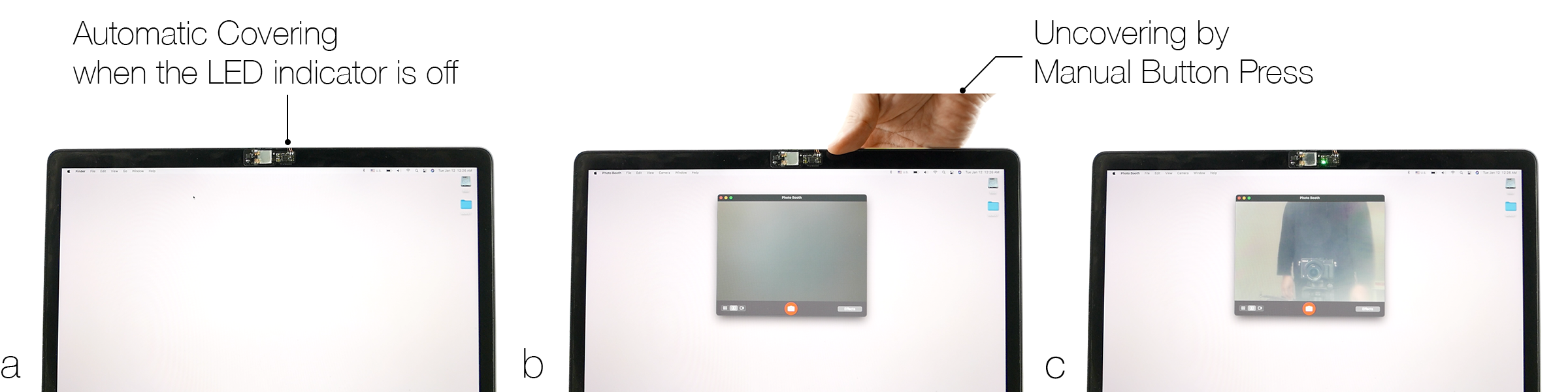}
    \caption{\swc employs automatic uncovering and manual covering for a webcam. (a) When end-users finish video applications, PDLC film of \swc turns opaque automatically, negating the need to remember to block a webcam. However, (b, c) Unlike covering, end-users are required to manually press a button of \swc, which makes the film turn transparent.  Adapted from Do et al. 2021 \cite{do2021smart}}
    \label{fig:swc}
\end{figure}

\subsubsection{\swc: Solution for Case Study 1 (Laptop Webcam)}
\swc (Figure~\ref{fig:swc}) is designed to automatically occlude a webcam when it is no longer in use. While users must manually uncover the webcam, \swc automatically deactivates itself when a laptop webcam's LED indicator turns off. The cover itself is made out of a polymer dispersed liquid crystal (PDLC) material, which is opaque by default but runs transparent when current flows through it. Accordingly, the material stays over the webcam regardless of its capture state.

\textbf{Direct physical manipulation of sensor state}: To uncover the webcam cover, users must manually press a button by the webcam cover. Pressing this button runs current through the PDLC material, making it transparent. In so doing, the user may access the webcam without occlusion. 

\textbf{Perceptible assurance of sensor state}: The cover remains over the webcam at all times, but is visibly opaque when deactivated and clear otherwise. Accordingly, the user can easily perceive the state of the sensor through a visual inspection.

\textbf{Intent-aligned Sensor (De)activation}: \swc detects the state of the laptop webcam indicator. If the webcam indicator is on, and the user presses the button to uncover the webcam, the PDLC material runs transparent to allow capture. This manual control of the cover requires the user to perform an intentional action to enable capture. However, when the webcam indicator is turned off, \swc automatically detects the change and turns opaque. This detection is done externally, through an air-gapped light detection sensor --- thus, attackers cannot suppress the LED indicator to covertly capture video feeds through the user's webcam, ensuring intent alignment.

\textbf{Impact on trust:} We ran a controlled, within-subjects experiment with 20 participants who are webcam cover users. We found that it improved trust---where trust was operationalized by users believing that the webcam would be covered when they wanted it covered---compared to manual webcam covers. Users trusted SWC more because they did not have to rely on their own memory to re-cover their webcams, and they could easily verify when the webcam cover was active.


\begin{figure}
    \centering
    \includegraphics[width=1\columnwidth]{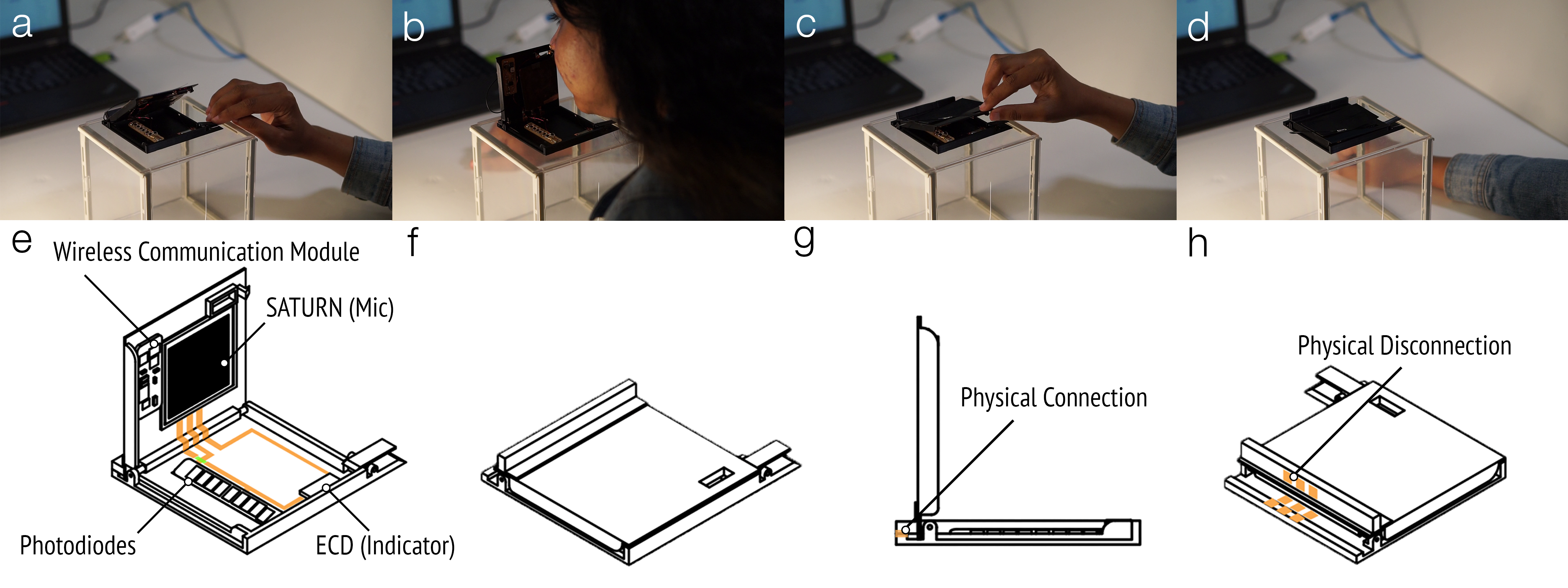}
    \caption{\cm is designed to expose its wiring between power modules and sensing and wireless communication modules. This allows visible power disconnection and connection based on users' intention. (a) End-users opens an clamshell casing manually. (b) Then, \cm is ready to record end-users' voice as the power module is connected at the hinge. (c, d) once finishing the voice recording, end-users can close the casing, which disconnects the power module at the hinge. The disconnection is physically visible, which provides perceptible assurance that the microphone cannot record unwittingly. Adapted from Do et al. 2023 \cite{do2023powering}}
    \label{fig:p4p}
\end{figure}

\subsubsection{\cm: Solution for Case Study 2 (Smart Speaker Microphone)}
\cm (Figure~\ref{fig:p4p}) is a wireless self-powered smart speaker microphone \cite{do2023powering}.
The key insight of \cm is to make whether or not the microphone is ``powered'' physically perceptible to end-users. 
Specifically, the wire connection of the power source, which consists of an array of photodiodes, is exposed.
All the electronics are embedded inside a clam-shell casing. When the casing is open, the photodiodes harvest energy and can power audio recording modules through the wire connected through the hinge of the casing. Otherwise, the power connection between the photodiodes and the audio recording electronics are disconnected, effectively disallowing capture.

\textbf{Direct physical manipulation of sensor state}: In order to activate the microphone, users must manually open the clamshell casing ensuring a link between intention and capture state. Users can also manually deactivate capture by closing the clamshell case.

\textbf{Perceptible assurance of sensor state}: \cm has a low-power indicator display that is physically-guaranteed to be magenta when there is no power, and green otherwise. The physical guarantee, again, comes the material the indicator is made out of---an electrochromic polymer material called ECP-Magenta. This material changes color from magenta to clear when a small, 0.45 V voltage is applied [ibid]. This material was placed on top of a green paper. When the clamshell is open and the indicator receives power, the ECP-magenta material runs clear and shows the green paper underneath. Otherwise, when there is no power, it remains magenta. Combined with the perceptible assurance of the clamshell state itself, because the shape of the clamshell is very different when open and when closed, \cm provides users with perceptible assurance of sensor capture state.

\textbf{Intent-aligned sensor (de)activation}: Since users must intentionally open the clamshell case when they want to be ``heard'' \cm also provides intent-aligned activation. It does not de-activate by default, however, suggesting an opportunity for future work to further enhance trust by employing a failsafe mechanism to deactivate the material.

\textbf{Impact on trust:} Through a controlled, within-subjects experiment with 16 participants who expressed privacy concerns against surreptitious recording by a smart speaker, \cm was found to improve trust compared to a commodity smart speaker. In this context, trust was operationalized as users' belief that \cm would not be able to capture audio when they did not explicitly want that audio captured. What drove this improved trust was the visibility of \cm's physical disconnection from its power source---participants had perceptible assurance that the microphone could draw no power when not in use. In contrast, participants had so much assurance using the mute button on a commodity smart speaker \cite{do2023powering}. 


\begin{figure}
    \centering
    \includegraphics[width=1\columnwidth]{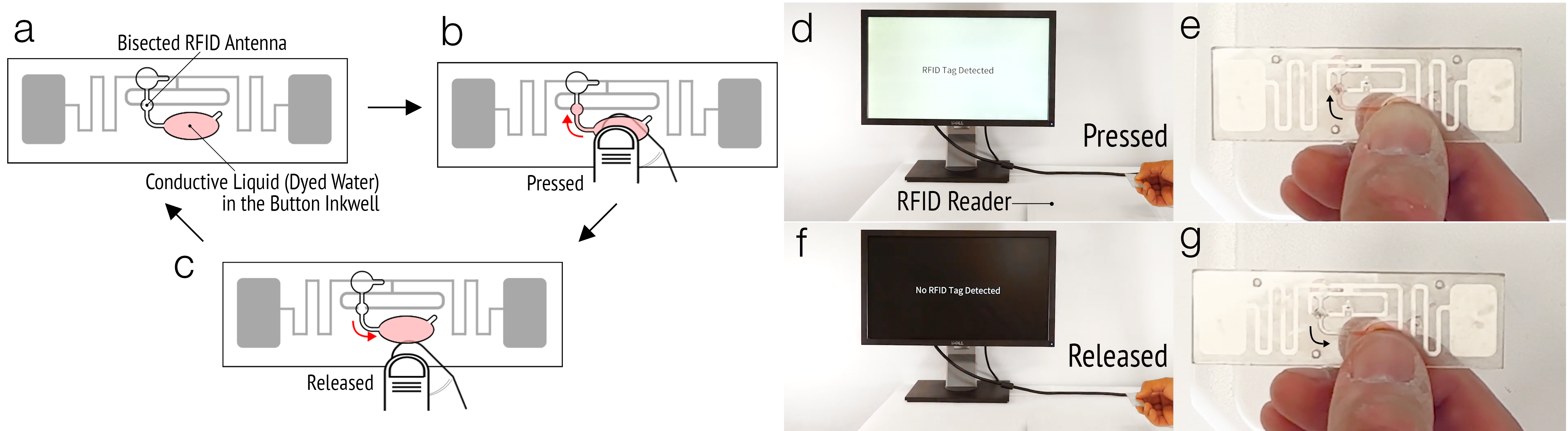}
    \caption{\odrfid allows end-users to make their RFID tags readable on demand. (a) By default, the antenna of the tag is disconnected. (b) When end-users intend to use the tag, they can press a button to push a visible ink stored in the tag, bridging the severed antenna and making the tag readable. (c) Once finishing the intent to use the tag, end-users can release their press, automatically retracting the ink and disconnecting the antenna. Adapted from Do et al. 2025 \cite{do2025demand}}
    \label{fig:odrfid}
\end{figure}

\subsubsection{\odrfid: Solution for Case Study 3 (Passive RFID Tag)}
On-demand RFID (Figure~\ref{fig:odrfid}) is an ``off-by-default'' passive RFID tag that makes it difficult, if not impossible, for adversaries to automatically sense its presence without a user's explicit knowledge and consent \cite{do2025demand}. \odrfid is implemented by integrating microfluidics technology \cite{sun2022microfluid,mor2020venous, wilson2022wearable}. Specifically, the antenna of \odrfid is bisected by default, disabling the RFID data transfer, similarly proposed as ``clipped tags'' by Karjoth and Moskowitz~\cite{karjoth2005disabling}. When a user wants to use the RFID tag, a user presses a well where conductive liquid is stored. This press pushes out the liquid to the bisected antenna, which connects the severed antenna trace and enables the RFID data transfer. Once the tag is not in need, a user can release their press, making the liquid automatically retreat to the well.

\textbf{Direct physical manipulation of sensor state}: In order to activate the RFID tag, the user must physically press down on the inkwell in the tag that contains the conductive liquid. Thus, the tag can only be activated through direct physical manipulation.

\textbf{Perceptible assurance of sensor state}: The conductive liquid ink in \odrfid is dyed red, so the user can see it traversing the microfluidic channel to reconnect the antenna when activated. When they let go, they can see that the liquid dye is concentrated in the inkwell, and that the antenna remains disconnected.

\textbf{Intent-aligned sensor (de)activation}: At the other end of the microfluidic channel in \odrfid is trapped air, which pushes the conductive ink back towards the inkwell when there is no oppositional force --- i.e., from a user pressing down on the inkwell. Thus, when the user releases their press, the liquid ink recedes back into the inkwell and disconnects the antenna again. Moreover,
the microfluidic channel in \odrfid is made out of a hydrophobic material, and thus little-to-no residual ink remains in the channel when the user is not actively pressing down on the inkwell. The combined effect is that \odrfid can only be activated when the user actively presses down on the inkwell.

\textbf{Impact on trust}: Through a within-subjects experiment with 17 participants, \odrfid was found to increase trust relative to commodity RFID tags both with and without RFID-blocking wallets. Trust in this context was operationalized as users' belief that a nearby RFID tag reader could not read their RFID tag unless the user explicitly wanted their tag to be read. Trust improved mainly due to two factors. First, \odrfid's antenna connection/disconnection status could be visually verified. Second, the physical mechanism through which a user's physical manipulation of the tag could result in the tag getting activated was clear to users because they could see the conductive ink traversing through the trace. 



\section{Discussion}

\psp in a new design paradigm for designing \snp controls for sensing-based systems can help build user trust that these systems are working in a manner aligned with their expectations.
We introduced three principles of \psp: (1) direct physical manipulation of sensor capture state; (ii) perceptible assurance of sensor state; and (iii) intent-aligned sensor (de)activation. Then, we highlighted the case studies to illustrate how \psp provides a path forward to rebuilding user trust in sensor-enabled systems. We next discuss its limitations, other complementary considerations that are important for building user trust in sensor-enabled systems, and outline a vision and agenda for future research.


\subsection{What about Sensors that One Does Not Own?}
As mentioned in \Cref{section:scope}, our focus thus far has been on sensors (e.g., cameras, microphones) or sensor accessories (e.g., RFID tags) that a user owns and can directly manipulate.
In these contexts, a primary user is actively choosing to use these sensor-based systems to unlock some benefits --- e.g., video conferencing, voice interfaces, and simple authentication --- but may be concerned that the privacy costs of these benefits are too high without additional assurances. 
In short, we explored physically-intuitive design principles for situations where users have an active choice as to whether or not to use the sensor-enabled system.
However, we have not yet discussed two situations that violate this assumption: institutional surveillance contexts, where users are subject to sensing from an institution (e.g., law enforcement or a workplace), and bystander privacy contexts, where users are subject to sensors installed by another individual. 

\subsubsection{Institutional Surveillance Contexts}
What we did not discuss is situations where users do not own or otherwise cannot manipulate a sensor / sensor accessory---e.g., surveillance cameras or adversarially placed sensors (except in the special case of passively sensed tags, like \odrfid). In these contexts, we argue it is not a reasonable design goal to ``build trust'': users are subjected to this surveillance, often without explicit consent and direct benefit. As such, we cannot expect to ``build trust'' by aligning sensor usage with user intention, and any interventions that are created with that goal will have to explore alternative multi-stakeholder approaches that resolve tensions between the surveilers and the surveilled. 



Note that we do expect that it may be possible to use physically-intuitive design principles to build \textit{resistance} interventions for users subject to surveillance.
For example, prior work has shown how it is possible to use facial masks and makeup to help users hide their faces from facial recognition algorithms \cite{monahan2015right}.
Exploring the design space of physically-intuitive resistance interventions to help users resist surveillance remains an interesting challenge to tackle.


\subsubsection{Bystander Privacy Contexts}

As the physical and digital worlds increasingly enmesh, so too are the social contexts in which individuals are situated. Increasingly often, people may find themselves in contexts where their data is being captured by sensors owned by other individuals in their periphery: e.g., when they enter a friend's ``smart home'', when they are going for a walk and encounter someone wearing smart glasses, or when in a shared office space.
In these contexts, too, as users are subject to sensors from which they do not directly benefit, it is not directly possible to ``build trust'' by aligning sensor capture with user intention.


A large body of prior work has focused on the problem of addressing bystanders' concerns against a primary user's device. Prior work emphasizes the importance of transparent communication, where device owners communicate how bystanders' data may be collected \cite{yao2019privacy,o2023privacy}. To that end, researchers have proposed various ways to improve communications about data collection to bystanders; these efforts include, e.g., creating accessible digital dashboards, and manipulating ambient lighting \cite{do2023vice, thakkar2022would}. Others have explored ways to provide bystanders with limited control over other users' sensors.
For instance, Steil et al. demonstrated head-mounted wearables that can close a physical shutter for a first-person camera based when in a sensitive environment \cite{steil2019privaceye}. Ahmad et al. proposed a new smart speaker that explicitly exposes microphone cable disconnection when muted, designed to clearly communicate the mute status to bystanders \cite{ahmad2022tangible}. 

However, these physical methods require a primary user to take proactive action on behalf of bystanders. To that end, prior work has highlighted the need to empower bystanders with agency over how their data is being captured by other user's devices \cite{yao2019privacy}. Prior work has also discussed why this is a thorny challenge. Thakkar et al., for example, found that users who own sensing-enabled devices often conceive of bystander privacy as a secondary concern \cite{thakkar2022would}. Others have discussed how mechanisms that enable a bystander to control others' devices may not be a realistic solution and cause undesirable social tension between a device owner and a bystander \cite{marky2020you}. Resolving these tensions will be challenging, but it is possible that physically-intuitive privacy and security design principles may provide a path forward: for example, if a bystander has some level of perceptible assurance that they are \textit{not} being captured, they may be more willing to trust the owner of a sensing-enabled device. 


\subsection{Complementary Design Considerations to Build Trust}

\subsubsection{Separating the Interests of the Sensor Manufacturer from Service Provider}
While \psp improves end-user trust in sensor-enabled devices, there remain challenges as trust is a multi-faceted construct. Prior work suggests that how end-users perceive device manufacturers is intertwined with how much they trust their devices \cite{jaspers2022consumers}, and distrust in device manufacturers could lead to non-use of their devices \cite{lau2018alexa}. Lau et al., for example, mentioned that smart speaker non-users believe that device manufacturers will not prioritize end-users interests over their own [ibid]. This finding suggests that some users may always harbor an implicit distrust of sensing-enabled devices as long as manufacturers have a vested interest in collecting more personal data. One way to address this concern in the future may be to explore mechanisms to separate the interests of the sensor manufacturers from the device manufacturers. Imagine if, for example, smart speakers did not come with sensors pre-installed but served instead as smart hubs that other sensors could easily connect to. Sensor manufacturers could then compete on privacy, usability, and other design values users may want to prioritize, while the smart speaker manufacturer simply interfaced with these trusted sensors to provide users with a desired service.


\subsubsection{Adversarial use}

Despite envisioning the positive intent of \psp, we must consider how to prevent harmful externalities if \psp design principles are widely adopted. Specifically, nefarious entities may find \psp as an opportunity for ``security theater'' \cite{schneier2003security} and ``privacy theater'' \cite{schwartz2008reviving}: i.e., ways to incorporate dramatic user interface operations where it \textit{seems} \snp has improved in spite of little improvement. Moreover, \snp operations could be designed to be seemingly `physically intuitive' but maliciously intended in a similar vein as malware. For example, with \cm, one could imagine having a super-capacitor in the circuit that allows the microphone to continue working even after the power is cut. This way, the effort to narrow the gap between how end-users think \snp operations work and how they actually function would end up being futile. There will remain challenges in how to apply \psp and to avert `theatrical' \snp operations.

\subsection{An Agenda for \psp Research}
The goal of \psp is to engender trust in smart, sensor-based systems. This trust, in turn, is rooted in bridging the gulf between how when users expect that their data is being captured and when it is actually being captured. We have shown, through three illustrative examples, that by following \psp design principles, it is possible to raise user trust in webcams, smart speaker microphones, and passive RFID tags.

But smart sensing environments are likely to comprise of many more sensors than the ones we explicitly covered in this paper. For example, the Mites sensor board---which was designed as a general-purpose sensing infrastructure for smart buildings---consists of ``nine discrete sensors with twelve unique sensor dimensions (vibration, thermal infrared, air pressure, magnetic field, light color, temperature, motion, Bluetooth devices, sound, WiFi signal strength, humidity, and light intensity).'' \cite{boovaraghavan2023mites} Moreover, one can imagine many other sensors in smart environments as well: depth cameras, touch sensors, force sensors, ultrasonic sensors, etc.

In addition to a variety of sensor setups, there is a spectrum of users with different levels of intuition in physicality. Delgado Rodriguez et al. found that different attributes (e.g., technology understanding and demographics) can affect their perceptions about different tangible privacy mechanisms \cite{delgado2024you}. This finding suggests the need of considering users' attributes to design \psp operations rather than finding one-size-fits-all solutions. 

Accordingly, there is a ripe opportunity to explore how we might employ \psp principles to promote trust in smart sensing environments for different users more broadly. We expect that \psp can be applied to a wide range of sensing-enabled systems where ensuring end-user trust is important for adoption and use. Furthermore, we envision \psp design principles could be adapted beyond a means to address \snp concerns against sensor \textit{data collection} to other parts of the data use pipeline, including \textit{data processing}, \textit{data storage}, and \textit{data dissemination}. 


For example, while not originally designed for \snp contexts, techniques for Functional Destruction also provide a hint that \psp design principles could be used for improving trust in data storage contexts \cite{cheng2023functional}. This work discusses various ways to produce electronics---including those that store data---that can be destroyed in an environmentally friendly manner. As with smart sensing systems, functions that promise to delete data also run the risk of being misaligned with user expectations. For example, when a user ``deletes'' their data from a digital system, they expect it to be destroyed. The reality is that traces of this data may remain and may be recoverable \cite{hughes2009disposal}.
Thus, physical destruction of devices is considered a secure way to ensure file deletion [ibid].
Functional Destruction techniques may provide users with a means to directly physically destroy their data, and to have perceptible assurance of data deletion by seeing that the hardware is destroyed.

In short, the design space for \psp is vast. This paper elucidates only the tip of the iceberg. 


\section{Conclusion}

In this paper, we introduce physically-intuitive privacy and security (\psp), a new design paradigm that increases user trust in sensor-enabled systems by designing \snp controls that take advantage of users' physical intuition.
Today, end-users harbor little trust in sensor-enabled systems: they believe, for example, that despite the presence of \snp controls, a smart speaker manufacturer can still eavesdrop on their conversations, and that third-party adversaries can remotely spy on users through their webcam.
Part of this distrust is fueled by the fact that \snp operations, to date, have not been designed in a manner that is intuitive: users have little ability to observe and verify how these controls actually work.
The key insight of \psp is to take advantage of people's intuitive understanding of how perception and mitigation in the physical world and apply them to \snp controls for sensor-enabled systems.
Doing so can increase trust by providing users with cleaner conceptual models of \textit{how} \snp controls work to enable or disable sensor capture.
Through an analysis of a series of research prototypes, we present three key principles to design systems for \psp---how to enable direct physical operations, understandable state changes, aligning sensor usage with users' intention of sensor usage.
We envision that consideration of these principles could usher in a future where users can trust in sensor-enabled systems, allowing, finally, for us to step closer towards Weiser's vision of the computer for the 21st century \cite{weiser1991computer}.



\bibliographystyle{ACM-Reference-Format}
\bibliography{references}

@article{mcknight2000trust,
  title={What is trust? A conceptual analysis and an interdisciplinary model},
  author={McKnight, D Harrison and Chervany, Norman L},
  year={2000}
}

@article{o2012general,
  title={A general definition of trust},
  author={O'Hara, Kieron},
  year={2012},
  publisher={University of Southampton}
}

@article{blomqvist1997many,
  title={The many faces of trust},
  author={Blomqvist, Kirsimarja},
  journal={Scandinavian journal of management},
  volume={13},
  number={3},
  pages={271--286},
  year={1997},
  publisher={Elsevier}
}

@article{weiser1991computer,
  title={The computer for the 21st century},
  author={Weiser, M},
  journal={Scientific American},
  pages={94--104},
  year={1991}
}

@article{abowd1998context,
  title={Context-awareness in wearable and ubiquitous computing},
  author={Abowd, Gregory D and Dey, Anind K and Orr, Robert and Brotherton, Jason},
  journal={Virtual Reality},
  volume={3},
  pages={200--211},
  year={1998},
  publisher={Springer}
}

@article{ahmad2020tangible,
  title={Tangible privacy: Towards user-centric sensor designs for bystander privacy},
  author={Ahmad, Imtiaz and Farzan, Rosta and Kapadia, Apu and Lee, Adam J},
  journal={Proceedings of the ACM on Human-Computer Interaction},
  volume={4},
  number={CSCW2},
  pages={1--28},
  year={2020},
  publisher={ACM New York, NY, USA}
}

@article{ahmad2022tangible,
  title={Tangible Privacy for Smart Voice Assistants: Bystanders' Perceptions of Physical Device Controls},
  author={Ahmad, Imtiaz and Akter, Taslima and Buher, Zachary and Farzan, Rosta and Kapadia, Apu and Lee, Adam J},
  journal={Proceedings of the ACM on Human-Computer Interaction},
  volume={6},
  number={CSCW2},
  pages={1--31},
  year={2022},
  publisher={ACM New York, NY, USA}
}

@article{do2021smart,
  title={Smart webcam cover: exploring the design of an intelligent webcam cover to improve usability and trust},
  author={Do, Youngwook and Park, Jung Wook and Wu, Yuxi and Basu, Avinandan and Zhang, Dingtian and Abowd, Gregory D and Das, Sauvik},
  journal={Proceedings of the ACM on Interactive, Mobile, Wearable and Ubiquitous Technologies},
  volume={5},
  number={4},
  pages={1--21},
  year={2021},
  publisher={ACM New York, NY, USA}
}

@inproceedings{do2023powering,
  title={Powering for privacy: improving user trust in smart speaker microphones with intentional powering and perceptible assurance},
  author={Do, Youngwook and Arora, Nivedita and Mirzazadeh, Ali and Moon, Injoo and Xu, Eryue and Zhang, Zhihan and Abowd, Gregory D and Das, Sauvik},
  booktitle={32nd USENIX Security Symposium (USENIX Security 23)},
  pages={2473--2490},
  year={2023}
}

@inproceedings{koelle2018beyond,
  title={Beyond LED status lights-design requirements of privacy notices for body-worn cameras},
  author={Koelle, Marion and Wolf, Katrin and Boll, Susanne},
  booktitle={Proceedings of the Twelfth International Conference on Tangible, Embedded, and Embodied Interaction},
  pages={177--187},
  year={2018}
}

@article{neustaedter2006blur,
  title={Blur filtration fails to preserve privacy for home-based video conferencing},
  author={Neustaedter, Carman and Greenberg, Saul and Boyle, Michael},
  journal={ACM Transactions on Computer-Human Interaction (TOCHI)},
  volume={13},
  number={1},
  pages={1--36},
  year={2006},
  publisher={ACM New York, NY, USA}
}

@inproceedings{machuletz2018webcam,
  title={Webcam covering as planned behavior},
  author={Machuletz, Dominique and Laube, Stefan and B{\"o}hme, Rainer},
  booktitle={Proceedings of the 2018 CHI Conference on Human Factors in Computing Systems},
  pages={1--13},
  year={2018}
}

@inproceedings{steil2019privaceye,
  title={Privaceye: privacy-preserving head-mounted eye tracking using egocentric scene image and eye movement features},
  author={Steil, Julian and Koelle, Marion and Heuten, Wilko and Boll, Susanne and Bulling, Andreas},
  booktitle={Proceedings of the 11th ACM symposium on eye tracking research \& applications},
  pages={1--10},
  year={2019}
}

@inproceedings{chen2020wearable,
  title={Wearable microphone jamming},
  author={Chen, Yuxin and Li, Huiying and Teng, Shan-Yuan and Nagels, Steven and Li, Zhijing and Lopes, Pedro and Zhao, Ben Y and Zheng, Haitao},
  booktitle={Proceedings of the 2020 chi conference on human factors in computing systems},
  pages={1--12},
  year={2020}
}

@inproceedings{sun2020alexa,
  title={" Alexa, stop spying on me!" speech privacy protection against voice assistants},
  author={Sun, Ke and Chen, Chen and Zhang, Xinyu},
  booktitle={Proceedings of the 18th conference on embedded networked sensor systems},
  pages={298--311},
  year={2020}
}

@inproceedings{truong2005preventing,
  title={Preventing camera recording by designing a capture-resistant environment},
  author={Truong, Khai N and Patel, Shwetak N and Summet, Jay W and Abowd, Gregory D},
  booktitle={UbiComp 2005: Ubiquitous Computing: 7th International Conference, UbiComp 2005, Tokyo, Japan, September 11-14, 2005. Proceedings 7},
  pages={73--86},
  year={2005},
  organization={Springer}
}

@inproceedings{brocker2014iseeyou,
  title={$\{$iSeeYou$\}$: Disabling the $\{$MacBook$\}$ Webcam Indicator $\{$LED$\}$},
  author={Brocker, Matthew and Checkoway, Stephen},
  booktitle={23rd USENIX Security Symposium (USENIX Security 14)},
  pages={337--352},
  year={2014}
}

@inproceedings{koscher2009epc,
  title={EPC RFID tag security weaknesses and defenses: passport cards, enhanced drivers licenses, and beyond},
  author={Koscher, Karl and Juels, Ari and Brajkovic, Vjekoslav and Kohno, Tadayoshi},
  booktitle={Proceedings of the 16th ACM conference on Computer and communications security},
  pages={33--42},
  year={2009}
}

@inproceedings{jin2022exploring,
  title={Exploring the needs of users for supporting privacy-protective behaviors in smart homes},
  author={Jin, Haojian and Guo, Boyuan and Roychoudhury, Rituparna and Yao, Yaxing and Kumar, Swarun and Agarwal, Yuvraj and Hong, Jason I},
  booktitle={Proceedings of the 2022 CHI Conference on Human Factors in Computing Systems},
  pages={1--19},
  year={2022}
}

@article{lau2018alexa,
  title={Alexa, are you listening? Privacy perceptions, concerns and privacy-seeking behaviors with smart speakers},
  author={Lau, Josephine and Zimmerman, Benjamin and Schaub, Florian},
  journal={Proceedings of the ACM on human-computer interaction},
  volume={2},
  number={CSCW},
  pages={1--31},
  year={2018},
  publisher={ACM New York, NY, USA}
}

@inproceedings{sciuto2018hey,
  title={" Hey Alexa, What's Up?" A Mixed-Methods Studies of In-Home Conversational Agent Usage},
  author={Sciuto, Alex and Saini, Arnita and Forlizzi, Jodi and Hong, Jason I},
  booktitle={Proceedings of the 2018 designing interactive systems conference},
  pages={857--868},
  year={2018}
}

@inproceedings{zeng2017end,
  title={End user security and privacy concerns with smart homes},
  author={Zeng, Eric and Mare, Shrirang and Roesner, Franziska},
  booktitle={thirteenth symposium on usable privacy and security (SOUPS 2017)},
  pages={65--80},
  year={2017}
}

@article{yao2019privacy,
  title={Privacy perceptions and designs of bystanders in smart homes},
  author={Yao, Yaxing and Basdeo, Justin Reed and Mcdonough, Oriana Rosata and Wang, Yang},
  journal={Proceedings of the ACM on Human-Computer Interaction},
  volume={3},
  number={CSCW},
  pages={1--24},
  year={2019},
  publisher={ACM New York, NY, USA}
}

@inproceedings{marky2020you,
  title={“You just can’t know about everything”: Privacy Perceptions of Smart Home Visitors},
  author={Marky, Karola and Prange, Sarah and Krell, Florian and M{\"u}hlh{\"a}user, Max and Alt, Florian},
  booktitle={Proceedings of the 19th International Conference on Mobile and Ubiquitous Multimedia},
  pages={83--95},
  year={2020}
}

@article{o2023privacy,
  title={Privacy-Enhancing Technology and Everyday Augmented Reality: Understanding Bystanders' Varying Needs for Awareness and Consent},
  author={O'Hagan, Joseph and Saeghe, Pejman and Gugenheimer, Jan and Medeiros, Daniel and Marky, Karola and Khamis, Mohamed and McGill, Mark},
  journal={Proceedings of the ACM on Interactive, Mobile, Wearable and Ubiquitous Technologies},
  volume={6},
  number={4},
  pages={1--35},
  year={2023},
  publisher={ACM New York, NY, USA}
}

@article{hutchins1985direct,
  title={Direct manipulation interfaces},
  author={Hutchins, Edwin L and Hollan, James D and Norman, Donald A},
  journal={Human--computer interaction},
  volume={1},
  number={4},
  pages={311--338},
  year={1985},
  publisher={Taylor \& Francis}
}

@inproceedings{ishii1997tangible,
  title={Tangible bits: towards seamless interfaces between people, bits and atoms},
  author={Ishii, Hiroshi and Ullmer, Brygg},
  booktitle={Proceedings of the ACM SIGCHI Conference on Human factors in computing systems},
  pages={234--241},
  year={1997}
}

@inproceedings{langheinrich2001privacy,
  title={Privacy by design—principles of privacy-aware ubiquitous systems},
  author={Langheinrich, Marc},
  booktitle={International conference on ubiquitous computing},
  pages={273--291},
  year={2001},
  organization={Springer}
}

@article{monahan2015right,
  title={The right to hide? Anti-surveillance camouflage and the aestheticization of resistance},
  author={Monahan, Torin},
  journal={Communication and Critical/Cultural Studies},
  volume={12},
  number={2},
  pages={159--178},
  year={2015},
  publisher={Taylor \& Francis}
}

@article{spiekermann2008engineering,
  title={Engineering privacy},
  author={Spiekermann, Sarah and Cranor, Lorrie Faith},
  journal={IEEE Transactions on software engineering},
  volume={35},
  number={1},
  pages={67--82},
  year={2008},
  publisher={IEEE}
}

@inproceedings{iravantchi2021privacymic,
  title={Privacymic: Utilizing inaudible frequencies for privacy preserving daily activity recognition},
  author={Iravantchi, Yasha and Ahuja, Karan and Goel, Mayank and Harrison, Chris and Sample, Alanson},
  booktitle={Proceedings of the 2021 CHI Conference on Human Factors in Computing Systems},
  pages={1--13},
  year={2021}
}

@inproceedings{naeini2017privacy,
  title={Privacy expectations and preferences in an $\{$IoT$\}$ world},
  author={Naeini, Pardis Emami and Bhagavatula, Sruti and Habib, Hana and Degeling, Martin and Bauer, Lujo and Cranor, Lorrie Faith and Sadeh, Norman},
  booktitle={Thirteenth Symposium on Usable Privacy and Security (SOUPS 2017)},
  pages={399--412},
  year={2017}
}

@article{dubois2020speakers,
  title={When speakers are all ears: Characterizing misactivations of iot smart speakers},
  author={Dubois, Daniel J and Kolcun, Roman and Mandalari, Anna Maria and Paracha, Muhammad Talha and Choffnes, David and Haddadi, Hamed},
  journal={Proceedings on Privacy Enhancing Technologies},
  volume={2020},
  number={4},
  year={2020}
}

@article{taylor2003most,
  title={Most people are “privacy pragmatists” who, while concerned about privacy, will sometimes trade it off for other benefits},
  author={Taylor, Humphrey},
  journal={The Harris Poll},
  volume={17},
  number={19},
  pages={44},
  year={2003},
  publisher={Harris Interactive Rochester, NY},
    howpublished = {\url{https://www.harrisinteractives.com/harris_poll/printerfriend-PID-365.html}},
    note = {(Accessed on 02/03/2024)}
}

@misc{HPSurvey71:online,
author = {Jenni Balthrop},
title = {HP Survey highlights webcam security and privacy behaviors},
howpublished = {\url{https://press.hp.com/us/en/press-releases/2019/awareness-of-webcam-hacking.html}},
month = {Jul},
year = {2019},
note = {(Accessed on 09/09/2020)}
}

@inproceedings{yao2019defending,
  title={Defending my castle: A co-design study of privacy mechanisms for smart homes},
  author={Yao, Yaxing and Basdeo, Justin Reed and Kaushik, Smirity and Wang, Yang},
  booktitle={Proceedings of the 2019 chi conference on human factors in computing systems},
  pages={1--12},
  year={2019}
}

@inproceedings{thakkar2022would,
  title={“It would probably turn into a social faux-pas”: Users’ and Bystanders’ Preferences of Privacy Awareness Mechanisms in Smart Homes},
  author={Thakkar, Parth Kirankumar and He, Shijing and Xu, Shiyu and Huang, Danny Yuxing and Yao, Yaxing},
  booktitle={Proceedings of the 2022 CHI Conference on Human Factors in Computing Systems},
  pages={1--13},
  year={2022}
}

@inproceedings{do2023vice,
  title={Vice VRsa: Balancing Bystander's and VR user’s Privacy through Awareness Cues Inside and Outside VR},
  author={Do, Youngwook and Brudy, Frederik and Fitzmaurice, George W and Anderson, Fraser},
  booktitle={Graphics Interface 2023-second deadline},
  year={2023}
}

@book{fitzmaurice1996graspable,
  title={Graspable user interfaces.},
  author={Fitzmaurice, George W and others},
  year={1996},
  publisher={University of Toronto, Department of Computer Science}
}

@article{cranor2008framework,
  title={A framework for reasoning about the human in the loop},
  author={Cranor, Lorrie F},
  year={2008},
  publisher={Advanced Computing Systems Professional and Technical Association}
}

@inproceedings{edwards2008security,
  title={Security automation considered harmful?},
  author={Edwards, W Keith and Poole, Erika Shehan and Stoll, Jennifer},
  booktitle={Proceedings of the 2007 Workshop on New Security Paradigms},
  pages={33--42},
  year={2008}
}

@article{sasse2001transforming,
  title={Transforming the ‘weakest link’—a human/computer interaction approach to usable and effective security},
  author={Sasse, Martina Angela and Brostoff, Sacha and Weirich, Dirk},
  journal={BT technology journal},
  volume={19},
  number={3},
  pages={122--131},
  year={2001},
  publisher={Springer}
}

@article{spiekermann2006technology,
  title={Technology paternalism--wider implications of ubiquitous computing},
  author={Spiekermann, Sarah and Pallas, Frank},
  journal={Poiesis \& praxis},
  volume={4},
  pages={6--18},
  year={2006},
  publisher={Springer}
}

@article{jaspers2022consumers,
  title={Consumers’ acceptance of domestic Internet-of-Things: The role of trust and privacy concerns},
  author={Jaspers, Esther DT and Pearson, Erika},
  journal={Journal of Business Research},
  volume={142},
  pages={255--265},
  year={2022},
  publisher={Elsevier}
}

@article{schonherr2020unacceptable,
  title={Unacceptable, where is my privacy? exploring accidental triggers of smart speakers},
  author={Sch{\"o}nherr, Lea and Golla, Maximilian and Eisenhofer, Thorsten and Wiele, Jan and Kolossa, Dorothea and Holz, Thorsten},
  journal={arXiv preprint arXiv:2008.00508},
  year={2020}
}

@inproceedings{vaidya2015cocaine,
  title={Cocaine noodles: exploiting the gap between human and machine speech recognition},
  author={Vaidya, Tavish and Zhang, Yuankai and Sherr, Micah and Shields, Clay},
  booktitle={9th $\{$USENIX$\}$ Workshop on Offensive Technologies ($\{$WOOT$\}$ 15)},
  year={2015}
}

@misc{bezos_2021, 
    title={Zeff Bezos - the Vanity Fair New Establishment Summit with Amazon CEO and Walter Isaacson.}, 
    publisher={YouTube}, 
    author={ACB English}, 
    year={2021}, 
    month={Sep},
    note = {\url{https://www.youtube.com/watch?v=5UGwFTdAk3I} (Accessed on 05/23/2023)},
}

@misc{smartspeakeruser_concerns,
  author = {NPR and Edison Research},
  title = {The Smart Audio Report},
  year = {2022},
  month = {June},
  note = {\url{https://www.nationalpublicmedia.com/uploads/2020/04/The-Smart-Audio-Report_Spring-2020.pdf} (Accessed on 01/31/2023)}
}

@inproceedings{egelman2010please,
  title={Please continue to hold},
  author={Egelman, Serge and Molnar, David and Christin, Nicolas and Acquisti, Alessandro and Herley, Cormac and Krishnamurthi, Shriram},
  booktitle={Ninth Workshop on the Economics of Information Security},
  year={2010}
}

@article{weiser1999origins,
  title={The origins of ubiquitous computing research at PARC in the late 1980s},
  author={Weiser, Mark and Gold, Rich and Brown, John Seely},
  journal={IBM systems journal},
  volume={38},
  number={4},
  pages={693--696},
  year={1999},
  publisher={IBM}
}

@inproceedings{madsen2000measuring,
  title={Measuring human-computer trust},
  author={Madsen, Maria and Gregor, Shirley},
  booktitle={11th australasian conference on information systems},
  volume={53},
  pages={6--8},
  year={2000},
  organization={Citeseer}
}

@article{seymour2023ignorance,
  title={Ignorance is bliss? the effect of explanations on perceptions of voice assistants},
  author={Seymour, William and Such, Jose},
  journal={Proceedings of the ACM on Human-Computer Interaction},
  volume={7},
  number={CSCW1},
  pages={1--24},
  year={2023},
  publisher={ACM New York, NY, USA}
}

@article{bambauer2013privacy,
  title={Privacy versus security},
  author={Bambauer, Derek E},
  journal={J. Crim. L. \& Criminology},
  volume={103},
  pages={667},
  year={2013},
  publisher={HeinOnline}
}

@inproceedings{medaglia2010overview,
  title={An overview of privacy and security issues in the internet of things},
  author={Medaglia, Carlo Maria and Serbanati, Alexandru},
  booktitle={The Internet of Things: 20 th Tyrrhenian Workshop on Digital Communications},
  pages={389--395},
  year={2010},
  organization={Springer}
}

@article{sylvester2005security,
  title={The security of our secrets: A history of privacy and confidentiality in law and statistical practice},
  author={Sylvester, Douglas J and Lohr, Sharon},
  journal={Denv. UL Rev.},
  volume={83},
  pages={147},
  year={2005},
  publisher={HeinOnline}
}

@book{norman2013design,
  title={The design of everyday things: Revised and expanded edition},
  author={Norman, Don},
  year={2013},
  publisher={Basic books}
}

@article{schneier2003security,
  title={Security Is a Weakest-Link Problem},
  author={Schneier, Bruce},
  journal={Beyond Fear: Thinking Sensibly About Security in an Uncertain World},
  pages={103--117},
  year={2003},
  publisher={Springer}
}

@article{schwartz2008reviving,
  title={Reviving telecommunications surveillance law},
  author={Schwartz, Paul M},
  journal={The University of Chicago Law Review},
  volume={75},
  number={1},
  pages={287--315},
  year={2008},
  publisher={JSTOR}
}

@article{sun2022microfluid,
  title={Microfluid: A multi-chip RFID tag for interaction sensing based on microfluidic switches},
  author={Sun, Wei and Chen, Yuwen and Chen, Yanjun and Zhang, Xiaopeng and Zhan, Simon and Li, Yixin and Wu, Jiecheng and Han, Teng and Mi, Haipeng and Wang, Jingxian and others},
  journal={Proceedings of the ACM on Interactive, Mobile, Wearable and Ubiquitous Technologies},
  volume={6},
  number={3},
  pages={1--23},
  year={2022},
  publisher={ACM New York, NY, USA}
}

@inproceedings{mor2020venous,
  title={Venous Materials: Towards Interactive Fluidic Mechanisms},
  author={Mor, Hila and Yu, Tianyu and Nakagaki, Ken and Miller, Benjamin Harvey and Jia, Yichen and Ishii, Hiroshi},
  booktitle={Proceedings of the 2020 CHI Conference on Human Factors in Computing Systems},
  pages={1--14},
  year={2020}
}

@article{wilson2022wearable,
  title={Wearable light sensors based on unique features of a natural biochrome},
  author={Wilson, Daniel J and Mart{\'\i}n-Mart{\'\i}nez, Francisco J and Deravi, Leila F},
  journal={ACS sensors},
  volume={7},
  number={2},
  pages={523--533},
  year={2022},
  publisher={ACS Publications}
}

@inproceedings{kelley2009nutrition,
  title={A" nutrition label" for privacy},
  author={Kelley, Patrick Gage and Bresee, Joanna and Cranor, Lorrie Faith and Reeder, Robert W},
  booktitle={Proceedings of the 5th Symposium on Usable Privacy and Security},
  pages={1--12},
  year={2009}
}

@inproceedings{windl2023investigating,
  title={Investigating tangible privacy-preserving mechanisms for future smart homes},
  author={Windl, Maximiliane and Schmidt, Albrecht and Feger, Sebastian S},
  booktitle={Proceedings of the 2023 CHI Conference on Human Factors in Computing Systems},
  pages={1--16},
  year={2023}
}

@inproceedings{cheng2023functional,
  title={Functional destruction: Utilizing sustainable materials’ physical transiency for electronics applications},
  author={Cheng, Tingyu and Tabb, Taylor and Park, Jung Wook and Gallo, Eric M and Maheshwari, Aditi and Abowd, Gregory D and Oh, Hyunjoo and Danielescu, Andreea},
  booktitle={Proceedings of the 2023 CHI Conference on Human Factors in Computing Systems},
  pages={1--16},
  year={2023}
}

@article{hughes2009disposal,
  title={Disposal of disk and tape data by secure sanitization},
  author={Hughes, Gordon F and Coughlin, Tom and Commins, Daniel M},
  journal={IEEE Security \& Privacy},
  volume={7},
  number={4},
  pages={29--34},
  year={2009},
  publisher={IEEE}
}

@incollection{zuboff2023age,
  title={The age of surveillance capitalism},
  author={Zuboff, Shoshana},
  booktitle={Social theory re-wired},
  pages={203--213},
  year={2023},
  publisher={Routledge}
}

@inproceedings{reeder2008expandable,
  title={Expandable grids for visualizing and authoring computer security policies},
  author={Reeder, Robert W and Bauer, Lujo and Cranor, Lorrie Faith and Reiter, Michael K and Bacon, Kelli and How, Keisha and Strong, Heather},
  booktitle={Proceedings of the SIGCHI Conference on Human Factors in Computing Systems},
  pages={1473--1482},
  year={2008}
}

@misc{defcon_2018, 
    title={DEF CON 26 - HuiYu and Qian - Breaking Smart Speakers We are Listening to You}, 
    publisher={YouTube}, 
    author={DEFCONConference}, 
    year={2018}, 
    month={Sep},
    note = {\url{https://youtu.be/3sLC0XaqvMg?feature=shared} (Accessed on 09/10/2024)},
}

@inproceedings{kroger2019my,
  title={Is my phone listening in? On the feasibility and detectability of mobile eavesdropping},
  author={Kr{\"o}ger, Jacob Leon and Raschke, Philip},
  booktitle={Data and Applications Security and Privacy XXXIII: 33rd Annual IFIP WG 11.3 Conference, DBSec 2019, Charleston, SC, USA, July 15--17, 2019, Proceedings 33},
  pages={102--120},
  year={2019},
  organization={Springer}
}

@article{boovaraghavan2023mites,
  title={Mites: Design and deployment of a general-purpose sensing infrastructure for buildings},
  author={Boovaraghavan, Sudershan and Chen, Chen and Maravi, Anurag and Czapik, Mike and Zhang, Yang and Harrison, Chris and Agarwal, Yuvraj},
  journal={Proceedings of the ACM on Interactive, Mobile, Wearable and Ubiquitous Technologies},
  volume={7},
  number={1},
  pages={1--32},
  year={2023},
  publisher={ACM New York, NY, USA}
}

@article{do2025demand,
  title={On-demand RFID: Improving Privacy, Security, and User Trust in RFID Activation through Physically-Intuitive Design},
  author={Do, Youngwook and Cheng, Tingyu and Wu, Yuxi and Oh, HyunJoo and Wilson, Daniel and Abowd, Gregory and Das, Sauvik},
  year={2025},
  publisher={Internet Society}
}

@inproceedings{karjoth2005disabling,
  title={Disabling RFID tags with visible confirmation: clipped tags are silenced},
  author={Karjoth, G{\"u}nter and Moskowitz, Paul A},
  booktitle={Proceedings of the 2005 ACM workshop on Privacy in the electronic society},
  pages={27--30},
  year={2005}
}

@inproceedings{chandrasekaran2021powercut,
  title={$\{$PowerCut$\}$ and obfuscator: An exploration of the design space for $\{$Privacy-Preserving$\}$ interventions for smart speakers},
  author={Chandrasekaran, Varun and Banerjee, Suman and Mutlu, Bilge and Fawaz, Kassem},
  booktitle={Seventeenth Symposium on Usable Privacy and Security (SOUPS 2021)},
  pages={535--552},
  year={2021}
}

@inproceedings{delgado2024you,
  title={Do you need to touch? Exploring correlations between personal attributes and preferences for tangible privacy mechanisms},
  author={Delgado Rodriguez, Sarah and Chatterjee, Priyasha and Dao Phuong, Anh and Alt, Florian and Marky, Karola},
  booktitle={Proceedings of the 2024 CHI Conference on Human Factors in Computing Systems},
  pages={1--23},
  year={2024}
}


\end{document}